\documentclass[]{tPHM2e}

\begin{document}
\doi{10.1080/14786435.20xx.xxxxxx}
\issn{1478-6443}
\issnp{1478-6435}
\jvol{00} \jnum{00} \jyear{2012} 

\markboth{H. Hodovanets et al.}{Philosophical Magazine}
\title{Boron isotope effect in single crystals of ErNi$_2$B$_2$C superconductor}

\author{Halyna Hodovanets, Sheng Ran, Paul C. Canfield, and Sergey L. Bud'ko $^{\ast}$\thanks{$^\ast$Corresponding author. Email: budkod@ameslab.gov \vspace{6pt}} \\\vspace{6pt}  {\em{Ames Laboratory, US DOE and Department of Physics \& Astronomy, Iowa State University, Ames, IA 50011, USA}}\\\vspace{6pt} }

\maketitle

\begin{abstract}
The influence of local moment magnetism on the boron isotope effect of $T_c$ was studied on single crystals of ErNi$_2$B$_2$C. Values of the partial isotope effect exponent of  $\alpha_B$=0.10$\pm$0.02 and $\alpha_B$=0.10$\pm$0.04 were obtained based on two different criteria applied to extract $T_c$. No significant change in the partial isotope effect exponent compared to the ones obtained for LuNi$_2$B$_2$C was observed. Based on this result we conclude that pair-breaking due to the Er local magnetic moment appears to have no detectable influence on boron isotope effect of $T_c$. \bigskip
\end{abstract}


\section{Introduction}
One of the accepted experimental probes of the extent to which superconductivity is phonon-mediated is the isotope effect \cite{fra94a,kis99a}. In the classical form of the Bardeen-Cooper-Schrieffer (BCS) theory, the isotope coefficient $\alpha$ (defined from $T_c \propto M^{-\alpha}$, where $M$ is the mass of the element) is equal to 1/2. More realistic theories consider slight deviation from $\alpha = 1/2$ \cite{mcm68a,car90a}. Measurements of even partial isotope effect (for one of the elements in the compound) were instrumental in addressing the mechanism of superconductivity in novel superconductors, e.g., non-magnetic borocarbides \cite{Lawrie,Cheon} and MgB$_2$ \cite{bud01a}.

A significant deviation of the isotope coefficient from the  BCS value was theoretically predicted for the case of superconductors containing noninteracting paramagnetic ions \cite{Abrikosov, Bill, Bill1, Kresin}. According to these theoretical views, the introduction of the magnetic scattering lowers $T_c$, and the increase of the pair-breaking scattering rate due to the paramagnetic impurities also increases the absolute value of the isotope effect exponent

\begin{equation}
\alpha=- \frac{d \ln(T_c)}{ d \ln(M)}= \frac{\alpha_0}{ 1-\rho \psi'(\rho+1/2)}
\end{equation}

where $\psi(x)$ is the digamma function, $\psi'(x)$ is the first derivative of the polygamma function with respect to its full argument, $\rho\equiv1/2\pi\tau_p(1+\lambda)k_B T_c$, $\tau_p$ is the scattering time that magnetic impurities are characterized by and $k_B$ is the Boltzmann constant \cite{Carbotte}. Here $\alpha _0$ is the isotope-effect exponent in the absence of the pair-breaking paramagnetic impurities. 

Figure 1 (after \cite{Carbotte}) schematically shows the normalized isotope effect exponent in the case of magnetic pair-breaking as a function of normalized temperature, $T_c$/$T_{c0}$, where $T_c$ and $T_{c0}$ are the critical temperatures of the superconductor with and without impurities. 

So far, the number of superconductors with paramagnetic impurities in which the isotope effect exponent was evaluated experimentally is rather small. In this work we report the measurements of the partial, boron, isotope effect exponent in single crystals of ErNi$_2$B$_2$C magnetic superconductor.

Since their discovery \cite{Nagarajan,Cava,Cava1}, the rare-earth nickel borocarbides, $\it R$Ni$_2$B$_2$C (where $\it R$=Y, Gd - Lu), have become a toy box for solid state physicists \cite{Canfield} due to the richness of physics one can find within the series, ranging from coexistence of magnetism and superconductivity to complex metamagnetism and to a classicall  heavy fermion behavior \cite{Canfield,mul01a,bud06a}.

$\it R$Ni$_2$B$_2$C crystallizes in a structure derived from the ThCr$_2$Si$_2$ structure, space group $\it I4/mmm$, that consists of alternating layers of Ni$_2$B$_2$ (in which Ni is tetrahedrally coordinated by four boron atoms) and sheets of RC \cite{Siegrist, Chakoumakos}. Phonon-mediated BCS superconductivity was proposed for borocarbides and supported, e.g., by specific heat measurements \cite{Carter}, however the anisotropy of the superconducting gap and the details of the superconducting pairing are still under discussion \cite{Maki, Yuan}. 

The phonons responsible for the superconductivity in these materials are thought to be the high-frequency boron A$_{1g}$ optical modes \cite{Mattheiss1, Hadjiev, Park, Bullock}. For non-magnetic, superconducting members of the series, the boron isotope effect (IE) of $T_c$ has been measured: significant boron IE was reported for polycrystalline YNi$_2$B$_2$C, partial isotope effect exponent, $\alpha_B$=0.25$\pm$0.04, \cite{Lawrie}; and for single crystals of YNi$_2$B$_2$C, $\alpha_B$=0.21$\pm$0.07, and LuNi$_2$B$_2$C, $\alpha_B$=0.11$\pm$0.05 \cite{Cheon}. The exponents for YNi$_2$B$_2$C and LuNi$_2$B$_2$C are different due to different molar masses, Debye temperatures and electron-phonon coupling constants of two compounds (Y has a molar mass of 88.9 g/mole and Lu has a molar mass of 175 g/mole; YNi$_2$B$_2$C has a molar mass of 240 g/mole and LuNi$_2$B$_2$C has a molar mass of 326 g/mole). 

\begin{figure}[t]
\begin{center}
{\resizebox*{10cm}{!}{\includegraphics{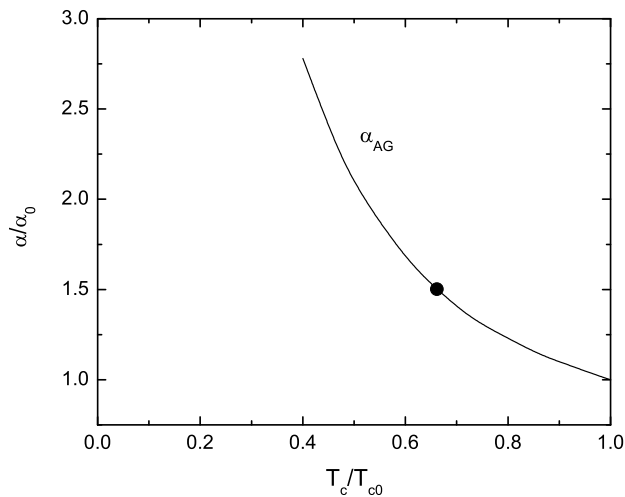}}}%
\caption{ Normalized isotope effect exponent $\alpha/\alpha_0$ vs $T_c/T_{c0}$ for Abrikosov-Gor'kov (AG) pair-breaking (after \cite{Carbotte}), $\alpha _0$ is the isotope-effect exponent in the absence of the pair-breaking. Filled point indicates expected value for ErNi$_2$B$_2$C under assumption of the AG pair breaking contribution.}
\label{fig:theory}
\end{center}
\end{figure}

Several $\it R$Ni$_2$B$_2$C compounds ($\it R$ = Dy - Tm) exhibit coexistence of superconductivity and magnetism. Our goal in this work is to explore whether the boron IE of $T_c$ is affected by the rare earth magnetism in the borocarbides. 

To check these predictions in the borocarbides family, boron IE measurements in lightly Gd-doped LuNi$_2$B$_2$C or YNi$_2$B$_2$C \cite{Bud'ko2} appear to be a suitable choice. Realistically though, with a very fast decrease of $T_c$ on Gd-doping, $dT_c/dx \approx 50$ K (that translates to $\sim 0.1$ K decrease in $T_c$ with 0.2\% of Gd), very precise control of Gd concentration in two batches (one with $^{10}$B, another with $^{11}$B) may be very difficult to achieve. 

For this work single crystals of stoichiometric ErNi$_2\,^{10}$B$_2$C and ErNi$_2\,^{11}$B$_2$C were chosen. The molar masses of Er and Lu are almost the same, therefore we expect Debye temperatures and phonon spectra of ErNi$_2$B$_2$C to LuNi$_2$B$_2$C to be similar as well. $T_c$ of ErNi$_2$B$_2$C is 10.5 K, and long-range antiferromagnetic order occurs below $T_N$=5.9 K \cite{Cho}; superconductivity appears in the paramagnetic state well above (by almost a factor of two) the antiferromagnetic ordering. In addition, the suppression of the superconducting transition temperatures for pure RNi$_2$B$_2$C (R = Ho - Tm) appear to follow the de Gennes scaling \cite{Canfield}, similar to what is expected for paramagnetic impurities. All of these observations point toward ErNi$_2$B$_2$C being an excellent test case for examining possible local moment corrections to the IE of $T_c$.

The ratio of $T_c$ of ErNi$_2$B$_2$C to that of LuNi$_2$B$_2$C is 0.66 that means that isotope effect exponent is expected to be $\sim 1.6$ times higher for ErNi$_2$B$_2$C than for LuNi$_2$B$_2$C if Eq. (1) is applied in this case. Given the error bars in $\alpha_B$ reported in ref. \cite{Cheon}, it appears to be challenging but still possible to distinguish between these two exponents. 
\section{Experimental methods}
Single crystals of ErNi$_2\,^{11}$B$_2$C and ErNi$_2\,^{10}$B$_2$C were grown using the flux growth technique \cite{Cho,Canfield5} with the only exception being that the ingots of ErNi$_2$B$_2$C and Ni$_2$B (flux) were arc-melted together in a ratio of 1:2.25 respectively (we found this ratio to give the best yield of crystals) before being placed in a vertical tube furnace. Same boron isotope was used for both ErNi$_2$B$_2$C and Ni$_2$B ingots. To ensure that extrinsic effects are not influencing $T_c$, the same furnace and the same temperature profile were used to grow samples containing $^{10}$B and $^{11}$B isotopes. The samples grow as plates with the $c$ axis perpendicular to the large flat surfaces of the samples. This single crystal growth technique has proven to be robust and reliable in the course of over a decade \cite{Cho,yar96a,cho01a,cre06a}.

To ensure sharp superconducting transitions \cite{Miao}, the samples were annealed. Nine samples of ErNi$_2\,^{11}$B$_2$C and nine samples of ErNi$_2\,^{10}$B$_2$C with the masses in a range from 2 to 25 mg were chosen for annealing. Each sample was wrapped into the Ta foil, then all samples, for each isotope separately, were placed into a fused silica tube that was continuously pumped on by the turbo-molecular pump to the pressure of $10^{-6}$-$10^{-7}$ Torr. The tube was heated to $950^0$ C initially for 100 h. The choice of the annealing temperature was based on Ref. \cite{Miao}. After the first 100 h, the furnace was turned off and the annealed samples were furnace cooled to the room temperature. Subsequently, the same samples were annealed for additional 200 h, following exactly the same procedure as a 100h annealing, and leading to a total annealing time of 300 h. Each sample was assign a label, from $\it a$ to $\it i$, and was tracked individually. This annealing yield the residual resistivity ratio values (for in-plane direction of the current) RRR=29 for ErNi$_2\,^{11}$B$_2$C and RRR=30 for ErNi$_2\,^{10}$B$_2$C. This improvement in RRR has been attributed to decreasing the number and density of dislocation lines in the crystal \cite{Avila}. Annealing also made the superconducting transition sharper. 

Magnetic measurements were carried out in a Quantum Design Magnetic Property Measurement System (MPMS) SQUID magnetometer with the field applied perpendicular to the $c$-axis. Zero-field cooled low field (25 Oe) magnetization data were taken on increasing temperature. The magnetization data were normalized by the absolute value of the diamagnetic moment at 4.6 K. The superconducting transition temperatures, inferred from magnetization data, were used to calculate the boron isotope constant $\alpha_B$. The criteria of extracting $T_c$ and calculation of boron isotope constant $\alpha_B$ will be explained in the next section.

\begin{figure}[t]
\begin{center}
{\resizebox*{10cm}{!}{\includegraphics{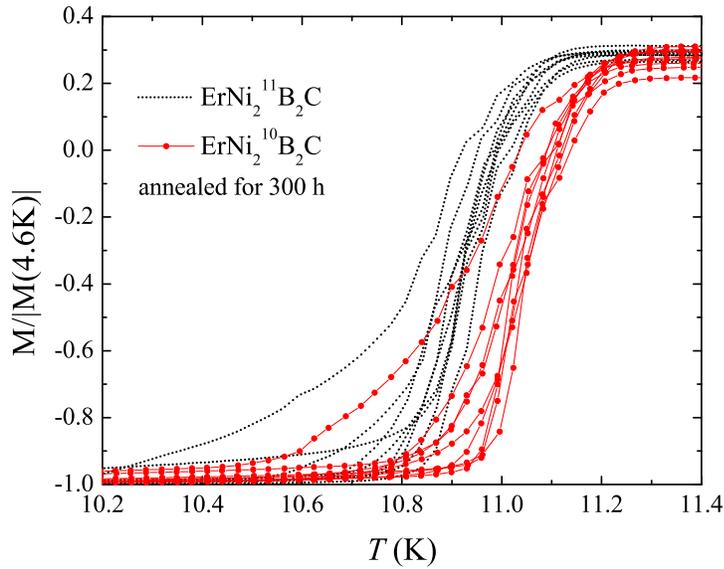}}}%
\caption{Temperature dependence of normalized magnetization data, M(T)/$|M(4.6K)|$, of single crystals of ErNi$_2\,^{11}$B$_2$C and ErNi$_2\,^{10}$B$_2$C annealed for 300 hours. Each curve comes from a different sample.}
\label{fig:Magnetization1}
\end{center}
\end{figure}

\begin{figure}[t]
\begin{center}
{\resizebox*{10cm}{!}{\includegraphics{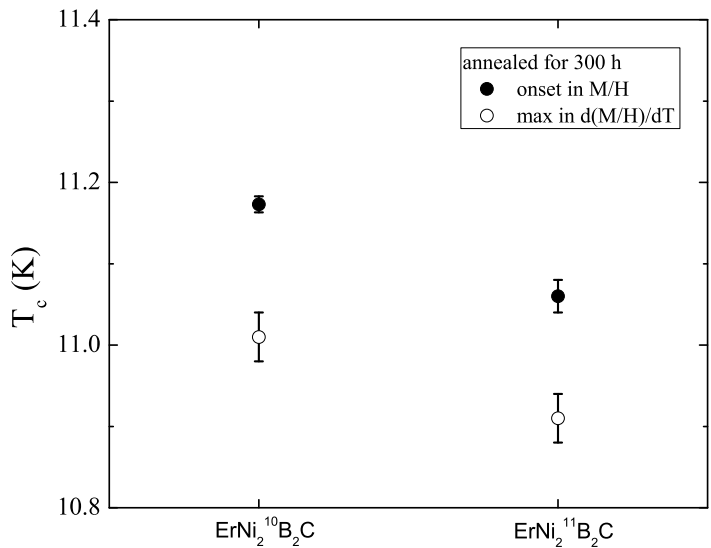}}}%
\caption{$T_c$ with standard deviation for ErNi$_2\,^{10}$B$_2$C and ErNi$_2\,^{11}$B$_2$C single crystals annealed for 300 hours.}
\label{fig:Tc}
\end{center}
\end{figure}

\section{Data and Discussion}
The normalized magnetization data for all measured single crystals of ErNi$_2\,^{11}$B$_2$C and ErNi$_2\,^{10}$B$_2$C annealed for 300 h are shown in Fig. 2. There is a small but observable isotope shift in the superconducting transition temperature $\Delta T_c^{IE}$=$T_c$($^{10}$B)-$T_c$($^{11}$B) between the data for ErNi$_2\,^{11}$B$_2$C and ErNi$_2\,^{10}$B$_2$C single crystals. 
To evaluate $\Delta T_c^{IE}$, the following two criteria for  $T_c$ determination were utilized: 1) onset in M/H and 2) maximum in $\frac{d(M/H)}{dT}$. These criteria allowed for the determination of the superconducting temperatures that are easily comparable among the samples. The data, with the standard deviation, for annealed for total of 300 h samples of ErNi$_2\,^{11}$B$_2$C and ErNi$_2\,^{10}$B$_2$C, are presented in Table 1. $T_c$, with standard deviation, versus annealing time is shown in Fig. 4 for better visualization. The inferred values of $\Delta T_c^{IE}$, 0.10$\pm$0.03 K and 0.10$\pm$0.04 K, are equal, within uncertainty, for both criteria and are smaller than the ones for LuNi$_2$B$_2$C (0.23$\pm$0.12 K and 0.16$\pm$0.07 K) \cite{Cheon}. 

To evaluate the possible statistical errors due to the measurement procedure and instrument the susceptibility of one sample of ErNi$_2\,^{10}$B$_2$C annealed for 300h was measured consequently nine times and $T_c$ of the sample was estimated using two criteria mentioned above. $T_c$ estimated from: 1) onset in M/H was (11.03$\pm$0.01) K and 2) maximum in $\frac{d(M/H)}{dT}$ was (11.15$\pm$0.004) K. This analysis indicates that the statistical errors are much smaller than the ones from the sample to sample scatter.

$T_c$ values given in Table 1 were used to calculate partial boron isotope effect given by equation (2)

\begin{equation}
\alpha_B = -\frac{\Delta \ln(T_c)}{\Delta \ln(M_B)},
\end{equation}

where $\Delta\ln(M_B)=\ln(\frac{10}{11})$ and $\Delta\ln(T_c)= \ln(\frac{T_c(^{10}B)}{T_c(^{11}B)})$ \cite{Lawrie}.

The partial boron isotope effect exponents for 300 h annealed samples are $\alpha_B$=0.10$\pm$0.02 and 0.10$\pm$0.04 for the two criteria mentioned above respectively. The partial boron isotope exponent calculated for the 300 h annealed samples of ErNi$_2$B$_2$C is equal, within uncertainty, to that of LuNi$_2$B$_2$C \cite{Cheon} despite Er being an ion with local magnetic moment. 

\begin{table}
  \tbl{Experimental results with standard deviation for ErNi$_2\,^{10}$B$_2$C and ErNi$_2\,^{11}$B$_2$C single crystals.}
{\begin{tabular}{@{}lcccccc}\toprule
   ErNi$_2$B$_2$C & $\#$ of samples & $T_c$ criteria & $\bar{T}_c$($^{10}$B)  (K) & $\bar{T}_c$($^{11}$B) (K) & $\Delta T_c^{IE}$ (K) & $\alpha_B$ \\
\colrule
   annealed 300 h & 9 with $^{10}B$, 9 with $^{11}B$ & onset in M/H &  $11.17\pm0.01$ & $11.06\pm0.02$ & $0.11\pm0.03$ & $0.10\pm0.02$ \\
   & & max in $\frac{d(M/H)}{dT}$ & $11.01\pm0.03$ & $10.91\pm0.03$ & $0.10\pm0.04$ & $0.10\pm0.04$ \\
   \botrule
  \end{tabular}}
\end{table}

The fact that Er bears local magnetic moment and introduces magnetic pair breaking into the system seems to have no detectable effect on boron isotope coefficient of $T_c$ which is different from the result of the theory that states that introduction of paramagnetic impurities increases the absolute value of isotope exponent\cite{Carbotte}. It has to be mentioned though that the theory was formulated for a very low concentration of the magnetic impurities, so that they can be considered non-interacting but the local magnetic moments on Er atoms interact via RKKY interaction that eventually is a cause of the long range magnetic order.  In addition, due to the crystal electric field effect the Er local magnetic moments are constrained to the $ab$ plane, that might affect superconductivity. More theoretical work needs to be done in order to determine how introduction of interaction among the paramagnetic impurities effects the absolute value of the isotope effect coefficient of $T_c$. 

\section{Conclusion}
The influence of the pair-breaking local magnetic moment in ErNi$_2$B$_2$C on the boron isotope effect was studied on the single crystals of ErNi$_2\,^{11}$B$_2$C and ErNi$_2\,^{10}$B$_2$C. Estimated $\Delta T_c^{IE}$=(0.11$\pm$0.03) K and (0.10$\pm$0.04) K lead to the partial boron isotope constant $\alpha_B$=0.10$\pm$0.02 and 0.10$\pm$0.04 respectively. The boron isotope constants obtained in this work are equal, within uncertainties, to that of LuNi$_2$B$_2$C \cite{Cheon} which suggests that the pair-breaking, despite lowering the superconducting temperature, appears to have no detectable effect on the boron isotope effect of $T_c$ in this system.

\section{Acknowledgments}
The authors would like to thank V.G. Kogan for fruitful discussions and reading of the manuscript. S. L. B. would like to thank Andreas Bill for bringing attention to this problem. The authors would also like to thank E. Colombier, R. Hu, H. Kim, S. Kim, X. Lin, E. D. Mun and A. Thaler for their support in writing this paper. This work was done at Ames Laboratory, US DOE, under contract $\#$DE-AC02-07CH111358. This work was supported by the US Department of Energy, Office of Basic Energy Science, Division of Materials Sciences and Engineering. S.L.B. and P.C.C were supported in part by the State of Iowa through the Iowa State University.

\end{document}